**Shaped angular dependence of the spin transfer torque and microwave generation without magnetic field**

O. Boulle[1], V. Cros[1], J. Grollier[1], L. G. Pereira[1,*], C. Deranlot[1], F. Petroff[1], G. Faini[2], J. Barnaś[3], A. Fert[1]

[1] Unité Mixte de Physique CNRS/Thales and Université Paris Sud XI, Route départementale 128, 91767 Palaiseau, France
[2] Laboratoire de Photonique et de Nanostructures LPN-CNRS, Route de Nozay, 91460 Marcoussis, France
[3] Department of Physics, Adam Mickiewicz University, Umultowska 85, 61-614 Poznań, Poland

Abstract: The generation of oscillations in the microwave frequency range is one of the most important applications expected from spintronics devices exploiting the spin transfer phenomenon. We report transport and microwave power measurements on specially designed nanopillars for which a non-standard angular dependence of the spin transfer torque (wavy variation) is predicted by theoretical models. We observe a new kind of current-induced dynamics that is characterized by large angle precessions in the absence of any applied field, as this is also predicted by simulation with such a wavy angular dependence of the torque. This type of non-standard nanopillars can represent an interesting way for the implementation of spin transfer oscillators since they are able to generate microwave oscillations without applied magnetic field. We also emphasize the theoretical implications of our results on the angular dependence of the torque.

The magnetization of a ferromagnetic body can be manipulated by transfer of spin angular momentum from a spin-polarized current. This is the concept of spin transfer introduced by Slonczewski [1] and Berger [2] in 1996. In most experiments, a spin-polarized current is injected from a spin polarizer into a "free" magnetic element, for example in pillar-shaped magnetic trilayers [3-6]. The phenomenon of spin transfer has a great potential for applications. It can be used either to switch a magnetic configuration (the configuration of a magnetic memory for example) [3-5] or to generate magnetic precessions and voltage oscillations in the microwave frequency range[6-7]. In the most usual situations, such oscillations are observed in the presence of a magnetic field.

From a fundamental point of view, spin transfer effects raise two different types of problems [8]. First the spin transfer torque acting on a magnetic element is related to the transverse spin polarisation of the current (transverse meaning perpendicular to the magnetization axis of the element) and can be derived from *spin-dependent transport equations* [8-17]. On the other hand, the description of the magnetic excitations generated by the spin transfer torque raises problems of *non-linear dynamics* [8,18-20]. For example, in the simple limit where the excitation is supposed to be a uniform precession of the magnetization (macrospin approximation), this precession can be determined by introducing the spin transfer torque into a Landau-Lifshitz-Gilbert (LLG) equation for the motion of the magnetic moment. However, the determination of the spin transfer torque and the description of the magnetization dynamics cannot be regarded as independent problems. In standard trilayered structures with in-plane magnetizations and with the usual angular dependence, a switching regime is found at zero and low magnetic field and the precession regime with generation of voltage oscillations is mainly observed above some threshold field [8]. We will show that a new behavior, characterized by large angle precessions in the absence of any magnetic field, can be obtained in specially designed structures presenting a non-standard dependence of the spin transfer torque as a function of the angle between the fixed magnetization of the polarizer and the magnetization of the free layer. This non-standard angular dependence of the torque, that we call "wavy", is obtained by choosing materials with different spin diffusion lengths for the "fixed" and "free" magnetic layers, which changes the distribution of the spin currents and spin accumulations in the structure.

The observation of spin transfer oscillations at zero field in structures with a "wavy" angular dependence of the torque can represent a new way to obtain spin transfer oscillators operating without any applied field, an other possible way being the use of exchange interactions or anisotropy to generate local effective fields or non-collinear equilibrium configurations [21]. In addition, the observation of a wavy angular dependence of the torque represents a valuable test of the theory and shows that realistic predictions of the spin transfer torque and its angular dependence in a given structure are now possible. As we will see, in the models we consider here [15-16], the torque is calculated from parameters which, for most of them, can be derived from former CPP-GMR experiments [22-23].

The usual behaviour observed in pillars with in-plane magnetizations along an anisotropy axis corresponds to the standard angular dependence of the inset of Fig.1a, in which the torque starts from zero at $\varphi = 0$ (P equilibrium state with parallel magnetizations of the fixed and free magnetic layers) and keeps the same sign till it comes back to zero at $\varphi = \pi$ (AP antiparallel state). At zero field and starting from a P state for example (Fig.1b), a negative current (electrons going from the free to the fixed layer in our convention) will destabilize the P state and stabilize the AP state, i.e. can switch the system from P to AP. In the presence of a large enough applied field favouring the P configuration, the torque cannot stabilize the AP state and leaves the system in an intermediate precession state. *This is what we call the standard behaviour with irreversible switching at low field and precession at high field, as illustrated by Fig.2a* (remark: in some low field experiments however, the irreversible switching is preceded by precessions in a very narrow current range just below the switching current).

The non-standard behaviour with precession at zero and/or low field presented in this article is related to the existence of a wavy angular dependence of the torque acting on the free magnetic layer. This oscillatory angular dependence, with an inversion of the torque between $\varphi = 0$ and $\varphi = \pi$, is shown in Fig. 1a. We present the results of calculations in the models of Fert *et al* [15] and Barnaś *et al* [16-17] for a Py(8)/Cu(10)/Co(8) pillar. With respect to standard structures like Co/Cu/Co or Py/Cu/Py, the difference we have introduced is a large asymmetry between the spin diffusion lengths (SDL) in the magnetic layers, with a long SDL in Co ≈ 38 nm (at room temperature) and a short SDL in Py ≈ 4 nm [22-23]. The smaller spin asymmetry of the resistivity in Co could also affect the angular dependence but we have checked by additional calculations that the wavy variation comes primarily from the shorter SDL in the Py free layer and not from the different spin asymmetry coefficients, as this has been mistakenly written in Ref.[24]. The solid curves in Fig.1a correspond to the calculation in the model of Barnaś *et al* [16]. A wavy angular dependence is also predicted by the model of Fert *et al* [15] which gives the terms of first order in $\varphi$ and $(\pi-\varphi)$ in the vicinity of the colinear P or AP states (the solid straight lines at the left and right edges of the graph in Fig.1). Due to the inversion at small values of $\varphi$, a negative current (Fig. 1c) now stabilizes not only the AP state but also the P one and should be 'inactive'. This can be a solution, for example, to reduce the spin-transfer-induced noise that is detrimental to read heads. In contrast, an appropriate positive current can destabilize both the P and AP states, leading to a precessional solution the motion equation, even at zero field.

To validate these predictions, we have performed transport and microwave power measurements at room temperature on Py(8)/Cu/Co(8) elliptical nanopillars of approximate dimensions 100x155 nm². Only the top Py layer (free layer) and the Cu spacer are etched through. The unetched Co layer ("fixed" layer) lies directly on the Ta/Cu bottom electrode. Very similar results have been obtained on Py(8)/Cu/Co(4)/IrMn nanopillars in which the extended Co layer is exchange biased by the IrMn one. We show in Fig. 2b the GMR signal of a Py(8)/Cu/Co(8) sample. Starting, for example, from large negative fields, the switching to an AP state at about 40 Oe is related to the magnetization reversal of the free layer (Py) to the positive direction, as this can be found from subsequent CIMS experiments in which the current-induced return to P is made harder by a larger positive field (consistently with a positive orientation of the Py magnetization in the switching to the AP state). From the GMR minor cycles of the Py layer (see Supplementary Information), we find that the coercive field of the Py layer is 90 Oe and the dipolar field acting on it is 43 Oe.

The different behaviours observed for standard and wavy angular dependences are first illustrated in Fig.2a and 2c. In Fig. 2a, we show the standard variation of differential resistance (dV/dI) versus I measured on a Py(4 nm)/Cu(10 nm)/Py(15 nm) pillar: starting from a P state, a negative current induces an irreversible switching from P to AP at low field and a reversible variation with the characteristic peak of steady precessions at high field. In contrast, starting again from a P magnetic configuration with magnetizations in the positive field direction but now with a Py(8 nm)/Cu(10 nm)/Co(8 nm) pillar for which a wavy angular dependence is expected, we detect (Fig. 2c) reversible peaks of dV/dI for positive currents and at very small fields on both sides of $H_{app}= 0$. The peak current increases with increasing applied positive field as expected since the P state becomes more stable. We have also performed experiments with an AP initial state. We find that dV/dI first drops to the level of a P state at some positive current and then, at higher current, exhibits the same characteristic precession peak we observe in measurements with a P initial state (data not presented).

In Fig. 3, we present microwave power spectra recorded with the same P initial state and for several values of the current. Fig.3a is for zero applied field (actually, $H_{app} \approx 2$ Oe) and Fig.3b for zero *effective* in-plane field (after subtracting the dipolar field). Coloured dots in the insets indicate the values of the current on the corresponding dV/dI vs I curves. A peak in the microwave power spectrum turns out approximately in some current range above the maximum of dV/dI. The frequency *f* of the microwave peak increases with the current (blue shift), in contrast with the red shift generally observed in standard pillars with in-plane magnetization. Actually, with the standard angular dependence of the torque, the theoretical prediction is a succession of red and blue shift regimes at increasing current but, in experiments with in-plane applied fields, the crossover to a blue shift regime has been seldom observed [25]. In macrospin simulations, a blue shift in *f* is predicted for the regime of out-of-plane (OP) precessions and is also associated with a decrease of *f* with increasing in-plane field. As shown in Fig.3 c, we observe this decrease of *f* with $H_{app}$. In Fig.4a, we present the current-field diagram of the microwave power. Microwave signals are emitted only in the top left corner of the diagram, i.e. at low field and in a zone which is also a region of increased resistance (Fig.4b). No excitation is observed at higher field.

We can therefore put forward two main results from our microwave power data: i) Pillars in which a wavy angular dependence of the spin transfer torque is expected, generate microwave oscillations, but, in contrast with the standard behavior, when there are excited by p*ositive* currents and *at zero field*; ii) These microwave oscillations present a blue shift of their frequency with current, a behaviour generally associated with out-of-plane precessions.

We first want to exclude that the effects described in the preceding paragraphs could arise from other origins than the wavy angular dependence of the STT. Could they arise from excitations of the Co "fixed" layer? We can first argue that the same behaviour is also observed when the 4nm thick and extended Co layer is pinned by an IrMn layer and that an excitation of a thin Co layer in the presence of such a strong pinning is quite improbable. We can also point out that, for un-pinned continuous magnetic layers, the switching current densities obtained by Chen *et al.* [26] are about one order of magnitude larger than ours. In addition, whereas a reduction of the thickness of the Co layer to 4 nm for the same 8nm Py thickness should make the excitation of Co easier (smaller current), our experimental results are in the opposite direction.

The sample of Fig.2-3 exhibits the relatively simple behaviour predicted for a wavy angular dependence of the torque in a macrospin picture, i.e. precessions at zero and low field in positive current. However, in a series of five similar samples (with or without pinning by IrMn), we have also observed additional features in transport measurements. For example, in some samples and with an initial P state, we see not only peaks in dV/dI in positive current at zero or low field but also partial or total switchings in negative current. These excitations can be ascribed to a non-uniform distribution of the magnetization [27]. For a part of the sample, the angle φ between the magnetizations of the two layers is above $φ_c$, the angle of torque inversion, and can be excited by a negative current. However, we emphasize that these additional excitations observed in transport measurements are never associated with peaks in the emitted power in the Gigahertz range. All the samples share the same main features with microwave emission only at low field in positive current.

We now present the theoretical implications of our experimental results and first comment briefly on the origin of the wavy angular dependence of the spin transfer torque in our samples. The physics governing this angular dependence can be discussed simply by considering that, in all the models [8,13-17] based on interfacial absorption of the transverse spin component and boundary conditions of the mixing conductance type (the language can be different in different formalisms), the spin transfer torque is proportional to the transverse component of the spin accumulation in the spacer layer. The key point is that the spin accumulation in a nonmagnetic conductor is directly related to the gradient of the spin current along the current axis z, $m \propto -d(j_m)/dz$ [28]. In configurations close to the P state of a standard pillar, with a thick fixed layer and a thin free layer made of the same material, the *spin polarization of the current in the spacer decreases from the fixed layer to the free layer*. This corresponds to a given sign of the spin accumulation. But an opposite sign is expected if, in the same configuration, the *spin polarization of the current increases from the fixed layer to the free layer*. This is what occurs for our Py(8nm)/Cu(10nm)/Co(8nm) pillars in an angular range close to the P configuration, as this can be seen from the spin accumulation calculated in the Section Methods. As shown in Fig 1a, calculations of STT based on two different models reflect this inversion of the spin accumulation by an inversion of the torque on the left part of the figure with respect to the standards case. However, as shown the figure, the inversion is a little less pronounced (less steep slope) in the model of Ref.[15] which goes beyond the simple mixing conductance approximation of Ref.[16].

For a further understanding, we have performed additional macrospin simulations of the current-induced precessions by solving a Landau-Lifschitz Gilbert equation including a spin transfer term using parameters compatible with the actual structure of the measured samples (see Methods, the simulations have been performed by two of the co-authors, O.B. and J.G., independently of those published in Ref.[24]). The simulated current-field diagram at T = 0 K is presented on Fig 4.d with a colour scale corresponding to the change of resistance. At high field ($H_{app}$ larger than the anisotropy field) and in the current range we have considered, the only excitations are in-plane (IP) precessions occurring above a threshold current $Ic_1$ and associated with a small change of resistance (which also corresponds to a small microwave power). At low field, the IP precessions above $Ic_1$ (black and blue trajectories in Fig. 4c) are followed by out-of-plane (OP) precessions (orange and red trajectories) above a second current threshold $Ic_2$.

There is a general good agreement between the main features of the experimental and calculated phase diagrams. In particular, the zone of OP precessions in the top left corner of the diagram of Fig. 4d turns out to be also the zone where we measure the larger DC resistance increase (Fig. 4b) and also detect microwave excitations (Fig.4a). Quantitatively, if one compares the colours in Fig.4 b and c, one can see that the distribution of the resistance change in the diagram is well reproduced and that the experimental ΔR in the OP zone is only somewhat smaller than the calculated one (by about 20% in average). The simulations also give a distribution of microwave power (not shown) concentrated in OP top-left zone as in the experimental plot of Fig.4a but with a power which is about 80 times larger than the experimental one. This could be due to several reasons. First, there are certainly technical factors, like a large impedance mismatch in the detection circuit. Second, for the OP excitations, the limits of a macrospin approach for a quantitative prediction [6,30], have been put forward by several publications. Finally, for the IP precessions we could not detect in the microwave spectra, it can be pointed out that a very small variation of GMR is expected for angles between P and an angle similar to our $\varphi_c$ in structures with our type of torque angular dependence [29]. This has also led us probably to overestimate the resistance change and the microwave power, since our calculation is based on a standard angular dependence of the GMR as $\sin^2(\varphi/2)$.

A confirmation that the zone of maximum resistance and microwave excitations in the top-left corner (positive currents and low fields) of the diagrams in Fig.4a-b can be identified with the zone of Out-of-Plane precessions in the calculated diagram (Fig. 4c) comes from the current and field dependence of the frequency. As shown in the inset of Fig. 4c, the simulations predict that a decrease of the frequency at increasing current for IP precessions is followed by an increase at the transition to OP precessions. This is in agreement with the frequency blue shift of the microwave excitations detected in the same zone of the phase diagram. The simulations also predict correctly the red shift for the variation with the field. Our simulations therefore support the picture of a non-standard behavior induced by a wavy angular dependence of the STT torque and characterized by out-of-plane precessions excited by positive current at zero and low field.

During the submission process, we learned that oscillations of vortex structures in thick Py layers excited by STT have been observed at relatively low field [31,32]. However this leads to oscillations at relatively low frequency, below 1 GHz for layers in our aspect ratio [33], and the oscillations above 3 GHz we observe cannot be explained by this mechanism.

Leaning on recent theoretical models of spin transfer torque, our experimental results should help designing more efficient spin transfer oscillators operating in a very small or even without an applied magnetic field. This is a necessary step (among others) on the implementation of these new spintronics-based oscillators in a microwave receiver system for telecommunication applications.

**Methods :**

The multilayers are grown by sputtering onto oxidized Si substrates. Two types of stacks were deposited : structure 1 = Au(20 nm)/Cu(5 nm)/Py(8 nm)/Cu(10 nm)/Co(8 nm)/Ta(10 nm)/Cu(80 nm)/Ta(10 nm) and structure 2 = Au(25 nm)/Py(8 nm)/ Cu(8 nm)/Co(4 nm)/IrMn(15 nm)/Ru(15 nm)/Cu(35 nm). Py stands for Permalloy. The results we present, are on a nanopillar with structure 1, but very similar results are observed with structure 2 when the fixed Co layer is pinned with an IrMn layer. This indicates that, even without an IrMn pinning layer, the magnetization of the extended Co layer is similarly fixed.

For the nanofabrication process, we first defined (by e-beam lithography, evaporation deposition and lift-off) a Ti(15 nm)/Au (55 nm) elliptical mask on the magnetic multilayer. Then, the magnetic pillar is etched by ion milling with a real-time monitoring by mass spectroscopy down to the Cu/Co interface. The bottom electrode is defined by optical lithography and ion milling. The next step is a planarization of the pillar with a Su-8 resist layer that is also used to electrically isolate the bottom and the top electrode. The Su-8 layer on the top of the pillar is removed by reactive ion etching. Finally, the top Ti/Au electrode is defined by optical lithography, evaporation deposition and lift-off.

We measured both the dc resistance and the differential resistance dV/dI using an additional 20μA ac current modulated at 5kHz. For the frequency-domain measurements, we applied a dc current on the sample through a bias-T. The high frequency voltage signal is then amplified (68 dB) and analysed on a commercial spectrum analyzer. The power spectra we show are extracted from the spectrum analyser (we do not correct them from a calibration done for quantities like the frequency-dependent amplifier gain, the attenuation in the transmission lines, and the impedance mismatches). They are only obtained by subtracting a reference spectra measured at *Idc* = 0 in the same magnetic field conditions. Note that the measured emitted power is therefore only a fraction of the actual emitted power from the pillars. Both transport and frequency measurements have been performed at room temperature and with in-plane magnetic field.

The torques of Fig.1a have been calculated by introducing in the models of Refs.[16] and [17] parameters mostly derived from CPP-GMR experimental data [22-23]. For respectively Au, Py, Cu, Co and Ta, these parameters are: bulk resistivity ρ (μΩ.cm) = 2, 15, 2.9, 24, 170; bulk spin asymmetry coefficient β = 0, 0.76, 0, 0.46, 0; spin diffusion length $l_{sf}$ (nm) = 35, 4, 350, 38, 10. For the interface parameters, respectively Au/Cu, Cu/Py, Cu/Co, Co/Ta, Ta/Cu, the parameters are : interfacial resistance $r_b$ (fΩ.m²) = 0.17, 0.5, 0.51, 0.5, 0.5; interfacial spin asymmetry coefficient γ : 0, 0.7, 0.77, 0.7, 0; interfacial spin memory loss coefficient δ : 0.13, 0.25, 0.25, 0.25, 0.1. Note that the values of the Co and Ta resistivity have been measured on thin film we had grown in the same conditions. The unknown value of $l_{sf}$ in Ta has been estimated by fitting the calculated and experimental variations of resistance ΔR. We have also used the same parameters in routine programmes [34] developed for the CPP-GMR to calculate the spin accumulation in the spacer layer for our structure and in a standard structure Py (15 nm)/Cu (10 nm)/Py(2 nm), respectively – 2.2 and + 2.0 in arbitrary units and check the change of sign at the origin of the wavy angular dependence.

For the simulations of the magnetization dynamics, we have solved a Landau Lifschitz Gilbert equation including a spin transfer term of the form $M_1 x(M_1 x M_2)$ with the angular dependence shown in Fig.1a. The calculations are performed at zero temperature. The saturation magnetization $\mu_0 Ms$ = 0.87 T has been derived from ferromagnetic resonance experiments performed on a Cu(6nm)/Py(7nm)/Cu(6nm) layer at room temperature. The other parameters are the anisotropy field $H_{an}$= 0.009 T, the gyromagnetic factor $\gamma_0 = 2.21\ 10^5 (s.A/m)^{-1}$, $\alpha = 0.011$. The area of the pillars is about $1.38\ 10^4$ nm², as derived by scanning electron microscopy.

**Correspondence** and requests for materials should be adressed to V. C. The authors declare they have no competing financial interests

**Acknowledgements**

The authors thank M. Gmitra for the calculations of Fig 1b based on the model of Ref[16]. We would like also to acknowledge H. Hurdequint for FMR measurements, L. Vila for assistance in fabrication, O. Copie and B. Marcilhac for assistance in transport and frequency measurements and M.R. Pufall for discussions. This work was partly supported by the french National Agency of Research ANR through the PNANO program (MAGICO PNANO-05-044-02) and the EU through the Marie Curie Training network SPINSWITCH (MRTN-CT-2006-035327). J. B. acknowledges support by funds from the Polish Ministry of Science and Higher Education as a research project (2006-2009).

*Present address : Instituto de Física, UFRGS, 91501-970 Porto Alegre, RS, Brazil

**Figure captions**

Figure 1 **Angular dependence of the spin transfer torque for a standard and a 'wavy' angular dependence. a,** Variation of the spin transfer torque on the free Py layer of a Au(infinite)/Cu(5 nm)/Py(8 nm)/Cu(10 nm)/Co(8 nm)/Ta(10 nm)/Cu(infinite) multilayer as a function of the angle φ between the magnetizations of the free Py and fixed Co layers for positive and negative currents. The solid curves are calculated in the model of Barnaś et al [17], the solid straight lines represent the slopes of the torque variation as the angle tends to 0 and π and have been derived from the small angle expression of Fert et al [16]. The parameters used in the calculations and mainly derived from CPP-GMR data are listed in the Section Methods. Inset : typical variation of the spin transfer torque as a function of the angle between the magnetizations of the free and fixed layers for a standard trilayer structure (case of Co/Cu/Co from Ref.[10]). **b-c,** Sketches showing schematically the direction (blue arrow) of the spin transfer torque on the free layer for configurations close to the P and AP configurations of the free layer (m) and fixed layer (M) magnetizations for a standard (b) and a wavy (**c**) angular dependence of the torque.

Figure 2 **Transport measurements on nanopillars with standard or "wavy" angular dependence of the spin transfer torque. a,** Differential resistance vs current measured for a nanopillar with a standard structure Py(15 nm)/Cu(10 nm)/Py(4 nm) at "low field" (H = 6 Oe) and "high field" (H = 133 Oe). In the latter case (precession), the applied field is larger than the coercive field equal to H = 133 Oe. Curves are offset for clarity. **b-c :** Transport data for a Co(8 nm)/Cu(10 nm)/Py(8 nm). nanopillar. **b,** Resistance vs field at low current (I = 200 µA). **c,** Differential resistance vs current for different applied fields around zero. These fields correspond to the coloured symbols in **b**.

Figure 3. **Microwave power spectra for the Co(8 nm)/Cu(10 nm)/Py(8 nm) nanopillar of Fig.2b-c. a,** Microwave power spectra for an applied field close to zero ($H_{appl}$ = 2 Oe) at different currents corresponding to the coloured symbols in the inset. Inset: dV/dI vs I for $H_{appl}$ = 2 Oe. **b,** Microwave spectra for different applied currents corresponding to the symbols in inset for an effective (applied + dipolar) field of about zero ($H_{app}$ = 43 Oe). Inset in **b** : dV/dI vs I for $H_{app}$ = 43 Oe. **c,** Microwave spectra for I = 9 mA at different positive applied fields. Spectra are offset for clarity.

Figure 4 **Experimental and simulated spin-transfer-induced high frequency dynamics for a Co(8nm)/Cu10nm)/Py(8nm) nanopillar**. **a,** Experimental integrated power between 0.1 to 8 GHz in colour scale as a function of field and current. **b,** Normalized experimental resistance in colour scale as a function of field and current (a reference curve has been subtracted to the experimental R vs I curves to remove the changes in resistance due to Joule heating). **c-d** : **Simulated dynamics of the magnetization in a macrospin approach c,** Results of macrospin numerical calculations of LLG equation as a function of current and field at T = 0K. The black line indicates the onset of current-induced precession. Inset in **c** : Variation of the calculated frequency as a function the current for $H_{app}$ = 0 Oe. **d** Magnetization trajectories for $H_{app}$=0 (black arrow in **c**) at several increasing applied currents.

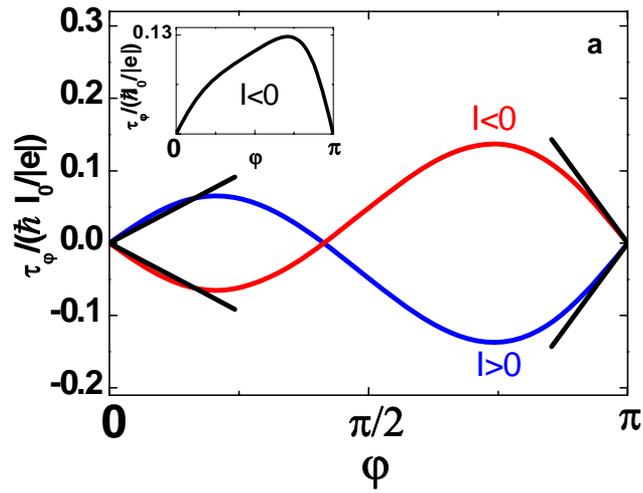
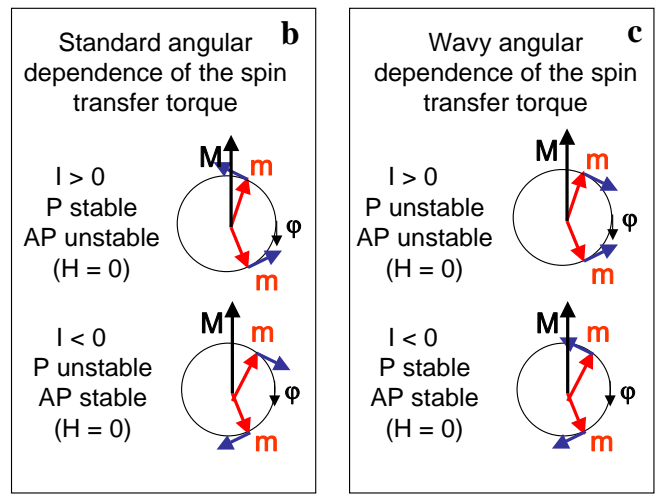

Fig. 1 Boulle *et al.*

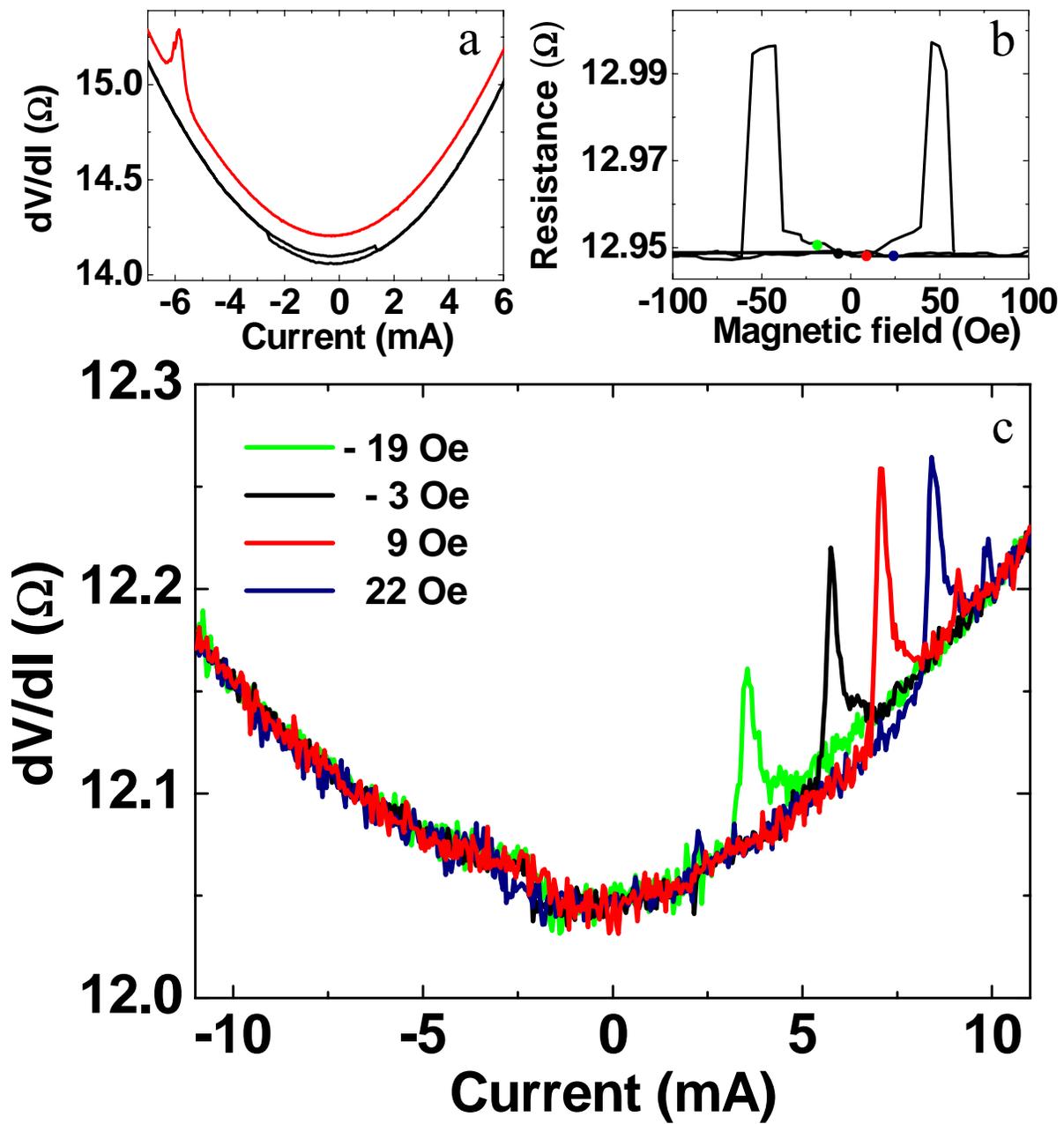

Fig. 2
Boulle *et al.*

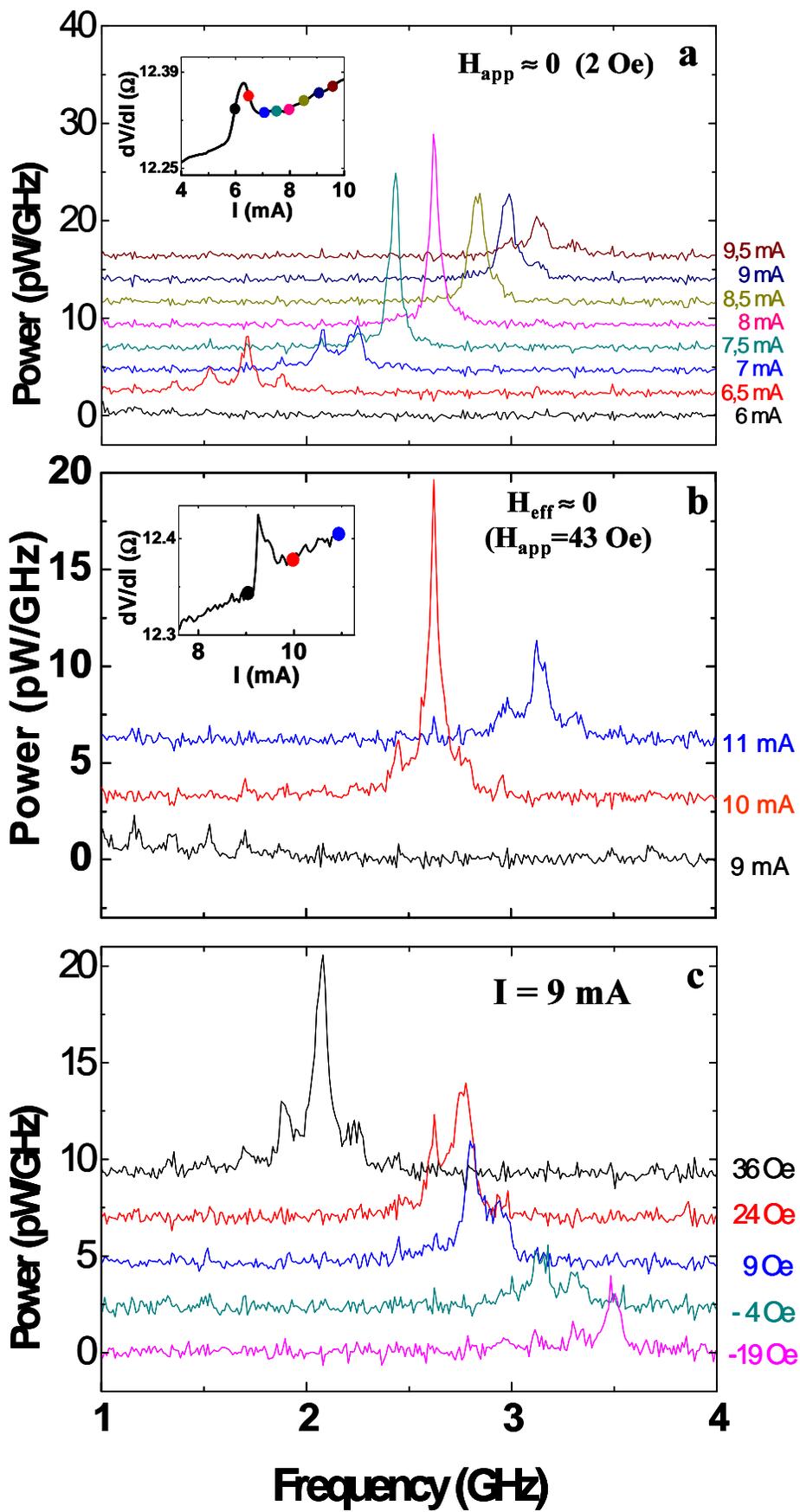

Fig. 3
Boulle *et al.*

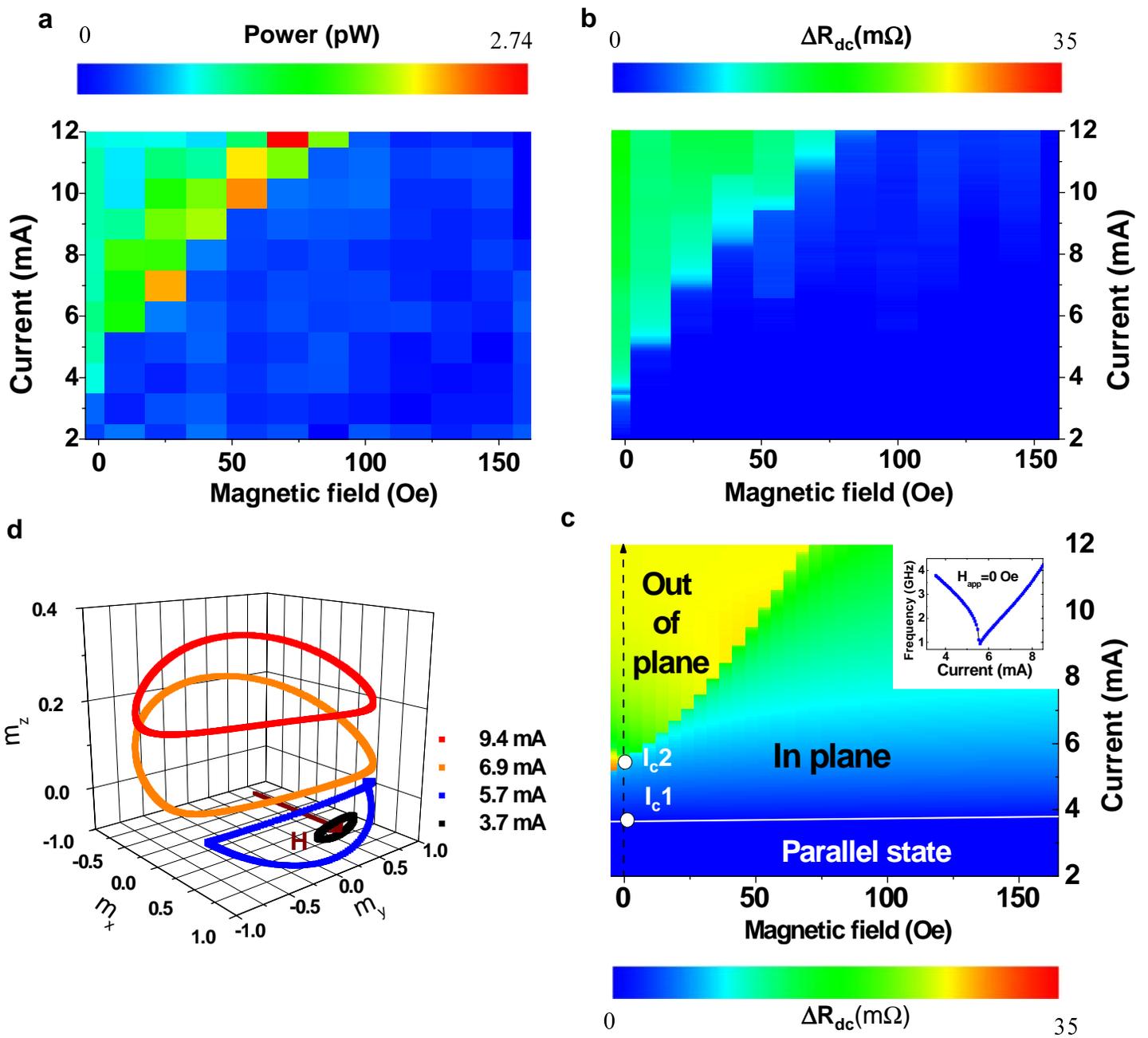

Fig 4. Boulle *et al.*